\documentclass{an}
\usepackage{graphicx}
\usepackage{times}
\Volume{00}              
\Year{0000}              
\Month{00}               
\Pagespan{000}{000}      

\begin{document}
\lhead[\thepage]{A.N. Author: Title}
\rhead[Astron. Nachr./AN~{\bf XXX} (200X) X]{\thepage}
\headnote{Astron. Nachr./AN {\bf 32X} (200X) X, XXX--XXX}

\title{Science Verification Results from PMAS}

\author{M.M. Roth, T. Becker, P. B\"ohm, A. Kelz}
\institute{Astrophysikalisches Institut Potsdam, An der Sternwarte 16,
D-14482 Potsdam, Germany}
\date{Received {date will be inserted by the editor}; 
accepted {date will be inserted by the editor}} 

\abstract{PMAS, the Potsdam Multi-Aperture Spectrophotometer, is a new
integral field instrument which was commissioned at the Calar Alto
3.5m Telescope in May 2001. We report on results obtained from a science
verification run in October 2001. We present observations of the low-metallicity
blue compact dwarf galaxy SBS0335-052, 
the ultra-luminous X-ray Source X-1 in the Holmberg$\;$II galaxy,
the quadruple gravitational lens system Q2237+0305 (the ``Einstein Cross''),
the Galactic planetary nebula NGC7027, and extragalactic planetary nebulae in
M31. PMAS is now available as a common user instrument at Calar Alto Observatory.
}
\correspondence{mmroth@aip.de}

\maketitle

\section{Introduction}

PMAS\footnote{http://www.aip.de/groups/opti/pmas/OptI\_pmas.html}
is a dedicated 3D instrument with a 16$\times$16 square element IFU
(0.5~arcsec pitch),
fiber-coupled to a fully refractive fiber spectrograph, which is based on
CaF$_2$ lenses and has good response in the blue. It is currently equipped 
with a 2K$\times$4K thinned CCD (SITe ST002A), providing 2048 spectral bins.
A 2$\times$2K$\times$4K mosaic CCD, which was commissioned 2003,
increases the free spectral range to 4096 spectral bins. 
The present fiber bundle has been conservatively manufactured with 100$\mu$m diameter, 
high OH$^-$ doped fibers for good UV transmission. A future upgrade with 50-60$\mu$m 
diameter fibers is intended to replace the existing IFU with a 32$\times$32 element array. 
A unique feature of PMAS is the internal A\&G camera, equipped with a LN$_2$-cooled,
blue-sensitive SITe TK1024 CCD, giving images with a scale of 0.2~arcsec/pixel
and a FOV of 3.4$\times$3.4~arcmin$^2$. The camera can be used with various broad-band
and narrow-band filters. For a more detailed description, see Roth et al.\ 2000a and 
Kelz et al.\ 2003. 
After First Light in May 2001, a Science Verification run was conducted at the
Calar Alto 3.5m Telescope in October 2001. Since then the instrument is available
at this telescope as a common user instrument. In this paper, we describe our
first results from the Science Verification observations. We selected targets with
well-known properties from the literature in order to assess whether PMAS is
capable of reproducing these data.

\begin{figure}[h]
\resizebox{\hsize}{!}
{\includegraphics[]{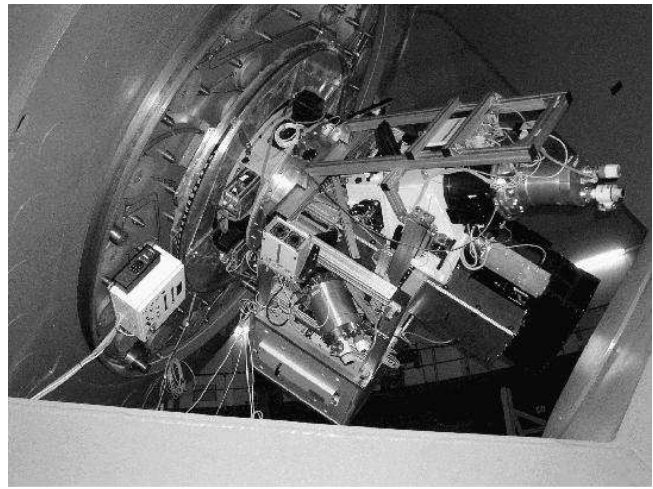}}
\caption{PMAS at the Cassegrain focus of the 3.5m Telescope at Calar Alto Observatory,
Spain.}
\label{PMAS}
\end{figure}

\section{SBS0335-052}

The blue compact dwarf galaxy SBS0335-052 is the second most metal-poor known galaxy 
after I Zw18, and thus an interesting target for spectrophotometric observations. 
Its oxygen abundance is 41 times lower than solar. It is thought to contain 6 embedded 
star clusters with  a significant number of supermassive stars of around 100 solar masses
(Thuan et al.\ 1997).
The intense far  UV radiation of those stars leads to high excitation ionization of the 
associated H$\;$II  regions, showing electron temperatures as high as 25000 K. 
The emission line spectrum of this galaxy has been studied sufficiently well in the 
literature, providing a good  test case for the PMAS science verification observations.

We observed SBS0335-052 on October 25, 2001, using the V600 grating, which yields a 
reciprocal dispersion of 0.8~{\AA}/pixel. In the 2$\times$2 binned readout mode of
the spectrograph CCD, which was used throughout this campaign, the dispersion was thus
1.6~{\AA}/bin. The spectral resolution was 3.3~{\AA} FWHM. 
The grating angle was set to cover
a wavelength range of 3600--5200~{\AA}. We took 7 exposures of 900sec each over a range
in airmass of 1.69$\ldots$1.35. The conditions were non-photometric with a seeing of
about 1.5arcsec FWHM.

\begin{figure}[h]
\resizebox{\hsize}{!}
{\includegraphics[]{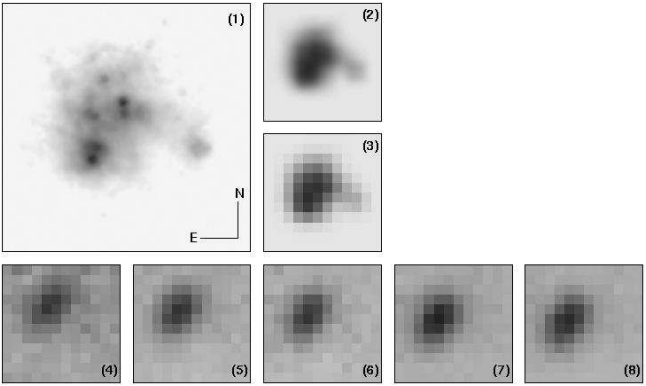}}
\caption{Direct images and 3D maps of SBS0335-052. Top: Thuan et al.\ 1997 HST image (1),
convolved with 1'' seeing (2), and resampled on a 0.5~arcsec grid (3). 
Bottom: 3D maps in [O$\;$II] 3727~{\AA} (4), continuum at 4200~{\AA} (5), 
H$_\gamma$ (6), H$_\beta$ (7), [O$\;$III] 4959~{\AA} (8).
Note the shifts with wavelength as an effect of atmospheric refraction.
}
\label{SBS0335-MAPS}
\end{figure}

\begin{figure}[h]
\resizebox{\hsize}{!}
{\includegraphics[]{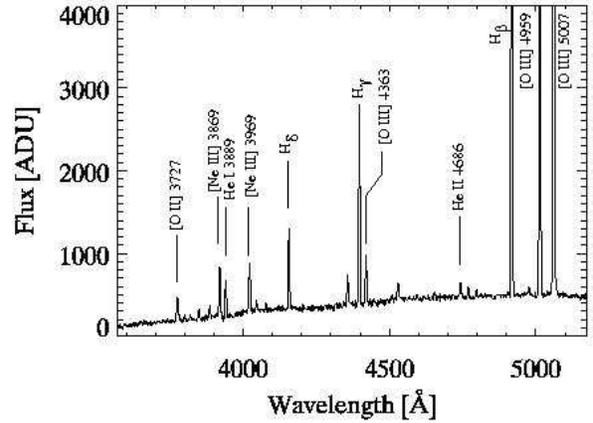}}
\caption{Co-added spectrum of SBS0335-052 from spaxels within the roughly elliptical region 
of high surface brightness in Fig.~\ref{SBS0335-MAPS}.}
\label{SBS0335-SPEC}
\end{figure}

Fig.~\ref{SBS0335-MAPS} shows the comparison of a direct HST WFPC2 image with several
monochromatic maps and the continuum.
The spectrum in Fig.~\ref{SBS0335-SPEC} was created from the combined data\-cube by 
defining a digital aperture and co-adding a total of 62 spatial elements 
(''spaxels'') within this aperture,
each measuring 0.5 x 0.5 arcsec squared. The roughly elliptical aperture was adapted to the 
elongated region of significant surface brightness along the NW direction.
The spectrum shows a rich emission line spectrum, including the Balmer series, 
the forbidden lines of [O$\;$III] 5007,4959
and [O$\;$III] 4363, [O$\;$II] 3727,3729, He$\;$I 3889, and He$\;$II 4686,
in good agreement with the spectra published by Melnick et al.\ (1992).

\section{Holmberg$\;$II X-1}

\begin{figure}[h]
\resizebox{\hsize}{!}
{\includegraphics[]{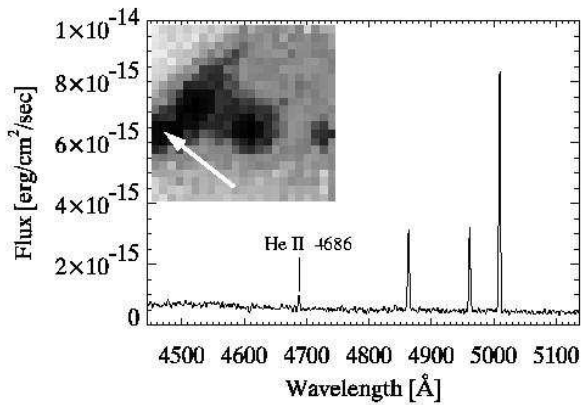}}
\caption{Co-added, flux-calibrated spectrum of a 2.5$\times$2.5~arcsec$^2$ aperture, 
confirming the presence of the He$\;$II 4686~{\AA} emission line. The insert shows a 
map in [O$\;$III] 5007~{\AA}, the arrow indicating the location of the He$\;$II emission line 
region. The map is a 11.0$\times$11.5~arcsec$^2$ mosaic of a total of 4 PMAS pointings
(North up, East left).}
\label{HoII}
\end{figure}

The ultraluminous X-ray source Holmberg$\;$II X-1
($\approx$10$^{40}$erg, assuming D=3.2 Mpc) is one of the X-ray brightest objects
in the vicinity of the Milky Way.  Zezas et al.\ (1999) have analyzed this source from
ROSAT PSPC and HRI data, finding a variable point source near the compact H$\;$II 
region \#70 (Hodge et al.\ 1994). 
\cite{pakull2002} have reported the discovery of high excitation He$\;$II 4686~{\AA}
emission which they attributed to the optical counterpart of HoII X-1. We observed
this object on October 28, 2001, using the V1200 grating, which provided a spectral 
resolution of 1.4~{\AA} FWHM and a wavelength coverage of 4450--5150~{\AA}. 
In accord with the results
of \cite{pakull2002}, we obtain maps in [O$\;$III] 5007~{\AA}, [O$\;$III] 4959~{\AA},
and H$_\beta$, and the spectrum is shown in Fig.~\ref{HoII}. The He$\;$II emission is constrained
to a region to the left of the insert map, coinciding with a Chandra
detection of Holmberg$\;$II X-1 within the error circle . The total flux in this emission line 
is 6$\times$10$^{-17}$erg/cm$^2$/sec. There is also a hint of a faint
continuum, increasing towards the blue. 
A more detailed analysis is presented by Lehmann et al. (in prep.).  

\section{Q2237+0305, the Einstein Cross}

The {\em Einstein Cross}, Q2237+0305, is a prominent example for a gravitational 
lens system of a distant quasar, showing a quadruple appearance with an image separation
of $\approx$2~arcsec. \cite{lewis1998} used CCD photometry and spectroscopy of 2 epochs,
separated by 3 years, to find evidence of microlensing from the foreground galaxy.
Images at the WHT were obtained with 0.5~arcsec FWHM seeing, revealing a brightening
of component A of 0.4~mag from 1991 to 1994. Data analysis of the spectroscopy
presented peculiar problems related to differential atmospheric dispersion, but
it was shown that the equivalent widths of C$\;$IV 1549~{\AA}, C$\;$III] 1909~{\AA}, 
and Mg$\;$II 2798~{\AA}  are substantially different bet\-ween
the four components, showing a drop by 30\% for the C$\;$IV line from 1991 to 1994.
\cite{mediavilla1998} observed this object with INTEGRAL at WHT under sub-arcsec seeing
conditions and reported the detection of an arc-like feature in C$\;$III] 1909~{\AA}.
\begin{figure}[h]
\resizebox{\hsize}{!}
{\includegraphics[]{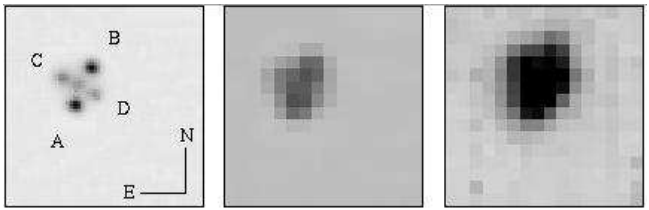}}
\caption{Direct WHT image of Q2237+0305 (left), degraded to spatial sampling of
0.5~arcsec (middle), PMAS map in C$\;$IV (right).}
\label{ECROSS}
\end{figure}
\begin{figure}[h]
\resizebox{\hsize}{!}
{\includegraphics[]{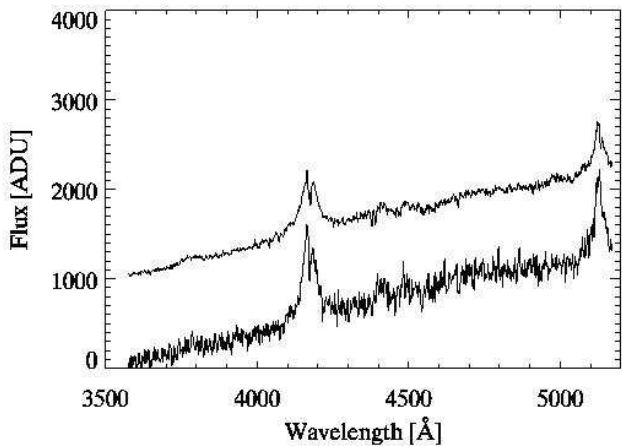}}
\caption{Spectra of components A and B in Q2237+0305.}
\label{ECROSS-SPEC}
\end{figure}

We observed Q2237+0305 on October 23 and 25, 2001, under less than ideal observing
conditions. Unfortunately, the seeing was no better than 1.3-1.5~arcsec, such that
the lens components were barely resolved. We used the V600 grating with a dispersion
of 1.6~{\AA}/bin, a spectral resolution of 3.3~{\AA} FWHM, and a wavelength range of 
3500--5200~{\AA}. We took a total of 6$\times$1800~sec exposures which turned out
to be of varying quality. 

In Fig.~\ref{ECROSS} and  Fig.~\ref{ECROSS-SPEC} we show 
the results of a single 1800~sec exposure, selected for the best seeing conditions 
among those 6 frames. A map in C$\;$IV 1549~{\AA} is shown in the right panel of
Fig.~\ref{ECROSS}, compared to the image provided by \cite{lewis1995} (left) and
the same image after convolution with 1~arcsec seeing and rebinning to the PMAS
sampling scale of 0.5~arcsec (middle).

The spectra for components A (upper plot) and B (lower plot)
were obtained by co-adding 3$\times$3 visually selected spaxels around the assumed
centroids with no further refinement of PSF fitting techniques. Sky correction was
performed by coadding 45 spaxels from an annulus around the target and subtracting
a scaled sky spectrum such that the sky lines vanished from the resulting spectrum.
Spectra A and B were scaled to match their continuum, but no correction for
instrumental response or flux calibration was applied. The plots in Fig.~\ref{ECROSS-SPEC}
are offset by 1000 ADU for clarity. As visible even to the eye,
the equivalent widths of the two components differ as observed by \cite{lewis1998}.
Prompted by this result, new PMAS observations were targetted at several new lens 
systems, see \cite{wisotzki2003}, and Wisotzki et al. (these proceedings).

\section{NGC7027}

NGC7027 is a bright Galactic planetary nebula with a very rich emission line
spectrum, and therefore an ideal target for instrumental tests and verification
observations. We consulted the catalogue of PN emission line intensities of \cite{kaler1976},
\cite{likkel1986}, and \cite{keyes1990} to compare our PMAS results with previously
published data.

We observed this object on October 24, 2001, under relatively favourable
seeing conditions of about 1~arcsec FWHM. The V600 grating was used (1.6~{\AA}/bin,
3.3~{\AA} FWHM resolution), and tilted to different angles, in order to cover the whole
optical wavelength range. Below we show a blue spectrum to obtain a qualitative
impression of the blue response, although at the chosen grating angle the efficiency
is way down the blaze function, and the choice of one of the UV gratings would have
been preferable for more serious work near the atmospheric cutoff. The exposure time
for this observation was 120~sec.
\begin{figure}[h]
\resizebox{\hsize}{!}
{\includegraphics[]{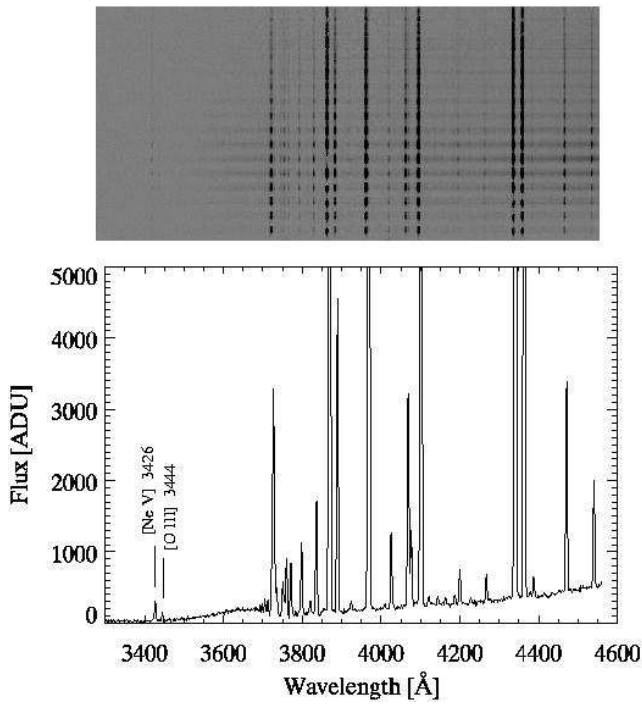}}
\caption{3D Spectra of the galactic planetary nebula NGC7027.}
\label{NGC7027-SPEC}
\end{figure}

The top panel of Fig.~\ref{NGC7027-SPEC} shows the fully reduced 256 spectra in a
stacked 2D format, which has some advantages for the purpose of inspection and
checking the data quality. The combined spectrum of 3$\times$3 
spaxels from the peak intensity region is plotted in the lower panel over the wavelength 
range of 3300--4550~{\AA}. Below 3400~{\AA} no useful signal was recorded.

The resulting spectrum shows a wealth of emission lines, superimposed on the very
bright nebular continuum. We do not further discuss the spectrophotometry of these
lines, but point out that [Ne$\;$V] and [O$\;$III] are visible at 3426~{\AA} and
3444~{\AA}, respectively, confirming that indeed the system has a non-negligible
response in the UV -- despite the less than ideal selection and setting of the grating.

\begin{figure}[h]
\resizebox{\hsize}{!}
{\includegraphics[]{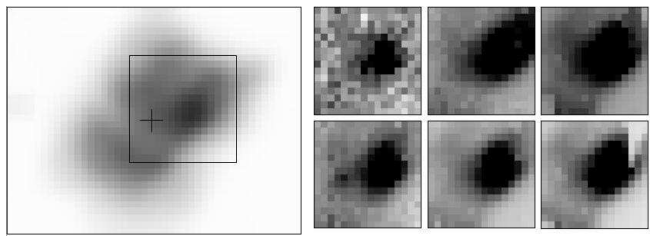}}
\caption{Direct image and 3D maps of NGC7027 (North up, East left).}
\label{NGC7027-MAPS}
\end{figure}

Fig.~\ref{NGC7027-MAPS} shows a comparison of a direct image of NGC7027 with
various pseudo-monochromatic maps, obtained from the datacube corresponding to
Fig.~\ref{NGC7027-SPEC}. The left frame shows the HST image published by
\cite{ciardullo1999}, rotated to the equatorial coordinate system and degraded to
roughly 1~arcsec resolution as one might expect it from the ground. The square
outlines the size of the 8$\times$8~arcsec$^2$ field-of-view of the PMAS IFU.
The cross near the center of the field indicates the position of the central
star, which is invisible in this degraded frame, but clearly stands out above the
background in the original HST image. The six frames to the right show PMAS 
maps at the following wavelengths (from left to right, top to bottom):
[Ne$\;$V] 3425, [O$\;$II] 3727, [O$\;$III] 4363, continuum 4500-4569~{\AA},
H$_\beta$, and H$_\gamma$ (the artifact in the H$_\gamma$ frame
is due to saturation). The [Ne$\;$V] map is, to our knowledge,
the first ground-based map obtained from a 3D datacube in the UV. Note that
the continuum image, being uncontaminated from any emission line, clearly 
reveals the presence of the central star, which has been known to be difficult 
to detect with direct imaging from the ground (Jacoby 1988).

\section{Planetary Nebulae in M31}

\begin{figure}
\resizebox{\hsize}{!}
{\includegraphics[]{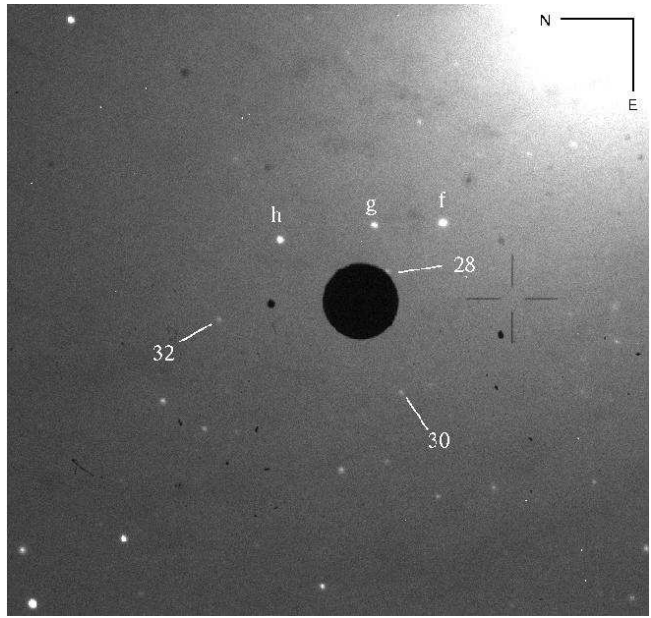}}
\caption{NE Field in M31, observed with internal A\&G CCD Camera.}
\label{CCD}
\end{figure}

PMAS was designed and built specifically with the goal to develop the
technique of {\em crowded field spectroscopy}, anaß-log\-ous to the now
well-established method of crowded field CCD photometry. These techniques
make extensive use of the property of {\em images} to exhibit a 
point-spread-function (PSF), which, if known, can be exploited for fitting
blended stellar images in severly crowded fields. 
As a pilot study, 
we have started at Calar Alto a guaranteed observing time project on 
the subject of extragalactic planetary nebulae (XPN) in the local group 
(Roth et al.\ in prep.). As a peculiar technical detail, Fig.~\ref{CCD}
shows the 3.4$\times$3.4~arcmin$^2$ FOV of the PMAS A\&G Camera, equipped
with a narrow-band [O III] filter. From a 2~min acquisition exposure, it is easily
possible to identify XPN of typically m$_{5007}\approx21$, which would otherwise
be impossible to detect with any standard video TV system.

\acknowledgements
We are indebted to G.\ Richter and the late V.\ Lipovetsky for paving our way into 
3D spectroscopy. We were kindly introduced into this fascinating field
by V.\ Afanasiev and S.\ Dodonov (Special Observatory for Astrophyics, Selentchuk).
The development of PMAS would not have become a reality without the help of the
ESO Optical Detector Team, allowing us to copy ESO-VLT detector head and CCD controller
designs. The extensive support of ESO is gratefully acknowledged. The PMAS Team would
like to thank Calar Alto staff for support during commissioning and operation. TB and AK 
acknowledge financial support of the German Verbundforschung under grants 05AL9BA1/9 
and 05AE2BAA/4.



\begin{thebibliography}{}


\bibitem[Ciardullo et al.\ (1999)]{ciardullo1999} Ciardullo, R. et al.\ 1999, AJ, 118, 488 

\bibitem[Hodge et al.(1994)]{hodge1994} Hodge, 
P., Strobel, N.~V., \& Kennicutt, R.~C.\ 1994, PASP, 106, 309 

\bibitem[Jacoby(1988)]{jacoby1988} Jacoby, G.~H.\ 1988, ApJ, 333, 
   193 

\bibitem[Kaler(1976)]{kaler1976} Kaler, J.~B.\ 1976, 
   ApJS, 31, 517 


\bibitem[Kelz, Roth, \& Becker (2003)]{kelz2003} Kelz, A.\ et al.\ 2003, SPIE, 4841, 1057

\bibitem[Keyes et al.\ (1990)]{keyes1990} Keyes, 
   C.~D., Aller, L.~H., \& Feibelman, W.~A.\ 1990, PASP, 102, 59 


\bibitem[Lewis et al.\ (1998)]{lewis1998} Lewis, G.F.\ et al.\ 1998, MNRAS, 295, 573

\bibitem[Likkel \& Aller (1986)]{likkel1986} Likkel, L.~\& Aller, 
L.~H.\ 1986, ApJ, 301, 825 

\bibitem[Mediavilla et al.\ (1998)]{mediavilla1998} Mediavilla, E.~et 
   al.\ 1998, ApJ, 503, L27 


\bibitem[Melnick et al.\ (1992)]{melnick1992} Melnick, J.\ et al.\ 1992, A\&A, 253, 16 


\bibitem[Pakull \& Mirioni (2002)]{pakull2002} Pakull, M.W., Mirioni, L. 2002,
   astro/ph 0202488

\bibitem[Roth et al.\ (2000)]{roth2000a} Roth, M.~M.~et al.\ 2000a,
   SPIE, 4008, 277

\bibitem[Thuan et al.\ (1997)]{thuan1997} Thuan, T.X., Izotov, Y.,
   Lipovetsky, V. 1997, ApJ, 477, 661

\bibitem[Lewis \& Irwin (1995)]{lewis1995} Lewis, G., Irwin,M. 1995,\\
http://antwrp.gsfc.nasa.gov/apod/ap950711.html


\bibitem[Wisotzki et al.\ (2003)]{wisotzki2003} Wisotzki, L.\ et al. 2003,
   A\&A 408, 455


\bibitem[Zezas et al.\ (1999)]{zezas1999} Zezas, A.~L.\ et al.\ 1999, MNRAS, 308, 302 



\end{thebibliography}
\end{document}